# On the Computation of Optimal Control Problems with Terminal Inequality Constraint via Variation Evolution

Sheng ZHANG, Yan-Qing CHENQ, and Wei-Qi QIAN

(2017.12)

*Abstract:* Studies regarding the computation of Optimal Control Problems (OCPs) with terminal inequality constraint, under the frame of the Variation Evolving Method (VEM), are carried out. The attributes of equality constraints and inequality constraints in the generalized optimization problem is traversed, and the intrinsic relations to the multipliers are uncovered. Upon these preliminaries, the right Evolution Partial Differential Equation (EPDE) is derived, and the costate-free optimality conditions are established. Besides the analytic expression for the costates in the classic treatment, they also reveal the analytic relations between the states, the controls and the (Lagrange and KKT) multipliers, which adjoin the terminal (equality and inequality) constraints. Moreover, in solving the transformed Initial-value Problems (IVPs) with common Ordinary Differential Equation (ODE) integration methods, the numerical soft barrier is proposed to eliminate the numerical error resulting from the suddenly triggered inequality constraint and it is shown to be effective.

## I. INTRODUCTION

Optimal control theory aims to determine the inputs to a dynamic system that optimize a specified performance index while satisfying constraints on the motion of the system. It is closely related to engineering and has been widely studied [1]. Because of the complexity, Optimal Control Problems (OCPs) are usually solved with numerical methods. Various numerical methods are developed and generally they are divided into two classes, namely, the direct methods and the indirect methods [2]. The direct methods discretize the control or/and state variables to obtain the Nonlinear Programming (NLP) problem, for example, the widely-used direct shooting method [3] and the classic collocation method [4]. These methods are easy to apply, whereas the results obtained are usually suboptimal [5], and the optimal may be infinitely approached. The indirect methods transform the OCP to a Boundary-value Problem (BVP) through the optimality conditions. Typical methods of this type include the well-known indirect shooting method [2] and the novel symplectic method [6]. Although be more precise, the indirect methods often suffer from the significant numerical difficulty due to the ill-conditioning of the Hamiltonian dynamics, that is, the stability of costates dynamics is adverse to that of the states dynamics [7]. The recent development, representatively the Pseudo-spectral (PS) method [8], blends the two types of methods, as it unifies the NLP and the BVP in a dualization view [9]. Such methods inherit the advantages of both types and blur their difference.

Theories in the control field often enlighten strategies for the optimal control computation, for example, the non-linear variable transformation to reduce the variables [10]. Recently, a new Variation Evolving Method (VEM), which is enlightened by the states evolution within the stable continuous-time dynamic system, is proposed for the optimal control computation [11]-[14]. The VEM

The authors are with the Computational Aerodynamics Institution, China Aerodynamics Research and Development Center, Mianyang, 621000, China. (e-mail: zszhangshengzs@hotmail.com).

also synthesizes the direct and indirect methods, but from a new standpoint. The Evolution Partial Differential Equation (EPDE), which describes the evolution of variables towards the optimal solution, is derived from the viewpoint of variation motion, and the optimality conditions will be asymptotically met under this frame. In Refs. [11] and [12], besides the states and the controls, the costates are also employed in developing the EPDE, and this increases the complexity of the computation. Ref. [13] proposed the compact version of the VEM that uses only the original variables, but it can only handles a class of OCPs with free terminal states. In Ref. [14], the compact VEM is further developed to address the OCPs with terminal Equality Constraint (EC). In this paper, we studied the situation that the terminal Inequality Constraint (IEC) is also included in the OCP formulation.

Throughout the paper, our work is built upon the assumption that the solution for the optimization problem exists. We do not describe the existing conditions for the purpose of brevity. Relevant researches such as the Filippov-Cesari theorem are documented in Ref. [15]. In the following, first the principle of the VEM is briefly reviewed. Then the attributes of ECs and IECs in the optimization problem are investigated as essential preliminaries. Next the VEM for OCPs including terminal IECs is studied. The evolution equations are derived, and the costate-free optimality conditions are established, which analytically relate the costates, the Lagrange multipliers and the Karush-Kuhn-Tucker (KKT) multipliers in the classic treatment to the state and control variables. Later illustrative examples are solved to verify the effectiveness of the method.

## II. PRINCIPLE OF VEM

The VEM is a newly developed method for the optimal solutions. It originates from the Lyapunov dynamics stability theory in the control field [16]. The system dynamics theory tells us that from stable dynamics, we may construct a monotonously decreasing function $V$, which will achieve its minimum when the equilibrium is reached.

For example, consider a continuous-time autonomous dynamic system like

$$\dot{x} = f(x) \tag{1}$$

where $x \in \mathbb{R}^n$ is the state, $\dot{x} = \dfrac{dx}{dt}$ is its time derivative, and $f : \mathbb{R}^n \to \mathbb{R}^n$ is a vector function. Suppose that $\hat{x}$ is a asymptotically stable equilibrium point of system (1) that satisfies $f(\hat{x}) = 0$. If $f(x)$ satisfies $(x - \hat{x})^\mathrm{T} f(x) < 0$ for any $x \neq \hat{x}$, then a feasible Lyapunov function can be constructed as

$$V = \frac{1}{2}(x - \hat{x})^\mathrm{T}(x - \hat{x}) \tag{2}$$

The dynamics governed by $f(x)$ determines that $\dot{V} \leq 0$. Thus $x$ will converge to the equilibrium $\hat{x}$ and $V$ will approaches the minimum of $V = 0$.

Inspired by it, now we consider its inverse problem, that is, from a performance index function (or functional) to derive the dynamics that minimize this performance index, and optimization problems are just the right platform for practice. Under this thought, the optimal solution is analogized to the stable equilibrium of a dynamic system and is anticipated to be obtained in an asymptotically evolving way. Since the OCPs seek the optimized variables, the fundamental Lyapunov theory, which aims to the dynamic system with finite-dimensional states, is accordingly generalized to the infinite-dimensional case as

**Lemma 1**: For an infinite-dimensional dynamic system described by

$$\frac{\delta y(x)}{\delta t} = f(y, x) \tag{3}$$

or presented equivalently in the Partial Differential Equation (PDE) form as

$$\frac{\partial y(x,t)}{\partial t} = f(y, x) \quad (4)$$

where "$\delta$" denotes the variation operator and "$\partial$" denotes the partial differential operator. $x \in \mathbb{R}$ is the independent variable, $y(x) \in \mathbb{R}^n(x)$ is the function vector of $x$, and $f : \mathbb{R}^n(x) \times \mathbb{R} \to \mathbb{R}^n(x)$ is a vector function. Let $\hat{y}(x)$, contained within a certain function set $\mathbb{D}(x)$, is an equilibrium function that satisfies $f(\hat{y}(x), x) = \mathbf{0}$. If there exists a continuously differentiable functional $V : \mathbb{D}(x) \to \mathbb{R}$ such that

i) $V(\hat{y}(x)) = c$ and $V(y(x)) > c$ in $\mathbb{D}(x)/\{\hat{y}(x)\}$.

ii) $\dot{V}(y(x)) \leq 0$ in $\mathbb{D}(x)$ and $\dot{V}(y(x)) < 0$ in $\mathbb{D}(x)/\{\hat{y}(x)\}$.

where $c$ is a constant. Then $y(x) = \hat{y}(x)$ is an asymptotically stable solution in $\mathbb{D}(x)$.

To implement the idea, a virtual dimension, the variation time $\tau$, is introduced to describe the process that a variable $x(t)$ evolves to the optimal solution to minimize the performance index within the dynamics governed by the variation dynamic evolution equations. Fig. 1 illustrates the variation evolution of variables in the VEM to solve the OCP. Through the variation motion, the initial guess of variables will evolve to the optimal solution.

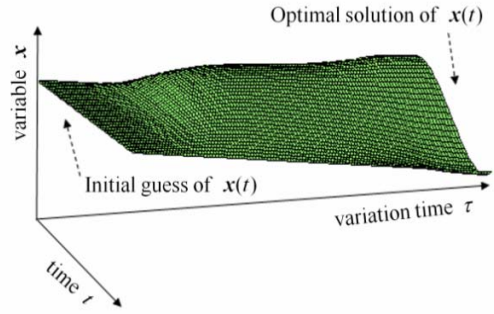

Fig. 1. The illustration of the variable evolving along the variation time $\tau$ in the VEM.

The VEM bred under this idea is first demonstrated for the unconstrained calculus-of-variations problems [11][13]. The variation dynamic evolution equations, derived under the frame of the VEM, may be reformulated as the EPDE and the Evolution Differential Equation (EDE), by replacing the variation operator "$\delta$" with the partial differential operator "$\partial$" and the differential operator "d". Since the right function of the EPDE only depends on the time $t$, it is suitable to be solved with the well-known semi-discrete method in the field of PDE numerical calculation [17]. Through the discretization along the normal time dimension, the EPDE is transformed to the finite-dimensional Initial-value Problem (IVP) to be solved, with the mature Ordinary Differential Equation (ODE) integration methods. Note that the resulting IVP is defined with respect to the variation time $\tau$, not the normal time $t$.

### III. PRELIMINARIES REGARDING ECS AND IECS

Because optimization problems with IECs are more intractable, preliminaries regarding the ECs and IECs, which help address the IECs in solving the OCPs with the VEM, are presented first. Consider the following generalized optimization problem formulation.

**Problem 1**: For the performance index

$$J_g = J_g(y(t), p) \quad (5)$$

subject to

$$C(y(t), p, t) \leq 0 \qquad t \in \mathbb{T}_I \qquad (6)$$

$$g(y(t), p, t) = 0 \qquad t \in \mathbb{T}_E \qquad (7)$$

where $t \in \mathbb{R}$ is the time, $y(t) \in \mathbb{R}^{n_y}(t)$ is the optimization variable vector and $p \in \mathbb{R}^{n_p}$ is the optimization parameter vector. Eq. (6) represents the IECs acting in the time set $\mathbb{T}_I$ and Eq. (7) refers to the ECs acting in the time set $\mathbb{T}_E$. Find the optimal solution $(\hat{y}(t), \hat{p})$ that minimizes $J_g$, i.e.

$$(\hat{y}(t), \hat{p}) = \arg\min(J_g) \qquad (8)$$

**Assumption 1**: For the optimal performance index, denoted by $\hat{J}_g$, the corresponding optimal solution $(\hat{y}(t), \hat{p})$ is unique.

*A. Positive-effect EC and Negative-effect EC*

**Definition 1**: Consider a specific time point $t_E \in \mathbb{T}_E$ and reformulate Eq. (7) as

$$g(y(t_E), p, t_E) = a \qquad (9)$$

where $a$ is a right dimensional vector. For the $i$ th component, if there is $\left.\dfrac{d\hat{J}_g}{da_i}\right|_{a_i=0} \leq 0$, then $g_i|_{t_E}$ is categorized as a Positive-effect EC (PEEC). In particular, $g_i|_{t_E}$ is named Pseudo-PEEC when $\left.\dfrac{d\hat{J}_g}{da_i}\right|_{a_i=0} = 0$; if $\left.\dfrac{d\hat{J}_g}{da_i}\right|_{a_i=0} > 0$, then $g_i|_{t_E}$ is categorized as a Negative-effect EC (NEEC).

**Theorem 1**: For a PEEC at a specific time point $t_E \in \mathbb{T}_E$

$$g_i(y(t_E), p, t_E) = 0 \qquad (10)$$

when in Problem 1 it is relaxed to be an IEC as

$$g_i(y(t_E), p, t_E) \leq 0 \qquad (11)$$

Presume the former optimal solution and the new optimal solution are located in the same basin (See Ref. [18]). Then we have

$$g_i(\hat{y}(t_E), \hat{p}, t_E) = 0 \qquad (12)$$

For a NEEC at $t_E$, when in Problem 1 it is relaxed as an IEC like Eq. (11), then under the same assumption we have

$$g_i(\hat{y}(t_E), \hat{p}, t_E) < 0 \qquad (13)$$

**Proof**: The IEC (11) may be equivalently re-presented as

$$g_i(y(t_E), p, t_E) - a_i = 0 \qquad (14)$$

where the slack number $a_i \leq 0$. According to Definition 1, we have $\left.\dfrac{d\hat{J}_g}{da_i}\right|_{a_i=0} \leq 0$. Note that $\dfrac{d\hat{J}_g}{da_i} \neq 0$ for $a_i \neq 0$ due to Assumption 1. Because the new optimal solution and the former optimal solution are in the same basin, to minimize $J_g$, there must be $a_i = 0$, or the optimal index $\hat{J}_g$ may decrease with $a_i < 0$. For the NEEC, Eq. (13) may be similarly proved. ∎

**Remark 1**: For the PEEC (10), when in Problem 1 it is relaxed to be an IEC as

$$g_i\left(\boldsymbol{y}(t_E), \boldsymbol{p}, t_E\right) \geq 0 \tag{15}$$

Presume the former optimal solution and the new optimal solution are located in the same basin. Then we have

$$g_i\left(\hat{\boldsymbol{y}}(t_E), \hat{\boldsymbol{p}}, t_E\right) \geq 0 \tag{16}$$

For a NEEC at $t_E$, when in Problem 1 it is relaxed as an IEC like Eq. (15), then under the same assumption we have

$$g_i\left(\hat{\boldsymbol{y}}(t_E), \hat{\boldsymbol{p}}, t_E\right) = 0 \tag{17}$$

**Remark 2**: Ignoring the PEEC (10) from Problem 1, and presuming the new optimal solution and the former optimal solution are in the same basin, we have

$$g_i\left(\hat{\boldsymbol{y}}(t_E), \hat{\boldsymbol{p}}, t_E\right) \geq 0 \tag{18}$$

For a NEEC at $t_E$, when in Problem 1 it is ignored, then under the same assumption we have

$$g_i\left(\hat{\boldsymbol{y}}(t_E), \hat{\boldsymbol{p}}, t_E\right) < 0 \tag{19}$$

*B. Active IEC and inactive IEC*

Now we consider the IEC (6) in Problem 1. Within the frame of an optimization problem, we classify the IECs according to their activeness at the optimal solution, that is

**Definition 2**: Consider a specific time point $t_I \in \mathbb{T}_I$, for an IEC

$$C_i\left(\boldsymbol{y}(t), \boldsymbol{p}, t_I\right) \leq 0 \tag{20}$$

it is said to be an active IEC if

$$C_i\left(\hat{\boldsymbol{y}}(t), \hat{\boldsymbol{p}}, t_I\right) = 0 \tag{21}$$

and it is said to be an inactive IEC if

$$C_i\left(\hat{\boldsymbol{y}}(t), \hat{\boldsymbol{p}}, t_I\right) < 0 \tag{22}$$

Note that an inactive IEC may be activated for some $\boldsymbol{y}(t)$ and $\boldsymbol{p}$ during the optimization process, but we will not call it an active IEC in this paper. From Definition 2, it is readily to find that

**Remark 3**: In Problem 1, strengthening an IEC (20) to be an EC as

$$C_i\left(\boldsymbol{y}(t), \boldsymbol{p}, t_I\right) = 0 \tag{23}$$

the optimal solution will not be changed if this IEC is an active IEC. Also, removing an inactive IEC from Problem 1, the optimal solution will not be changed.

**Theorem 2**: The IEC (20) is an active IEC if and only if the strengthened EC (23) is a PEEC, and the IEC (20) is an inactive IEC if and only if EC (23) is a NEEC.

**Proof**: we only prove the first statement of this theorem, because the proof for the second part is similar. Regarding the necessity, if EC (23) is a PEEC, then from Theorem 1, its relaxed IEC (i.e., Eq. (20)) is an active IEC from Definition 2. We use reduction to absurdity to show the sufficiency. If the strengthened EC (23) is a NEEC, then from Theorem 1, we know that IEC (20) will be an inactive IEC, which contradicts with the premise that the IEC (20) is an active IEC. ∎

*C. Relations to the multipliers*

Theorem 2 shows the intrinsic relations between the IECs and the ECs, and the IECs may be distinguished from their strengthened ECs. However, in practices it will not be an easy task to determine a PEEC or a NEEC from its definition. Fortunately, the well-known multipliers, which are usually employed for the optimality conditions, imply their type.

**Theorem 3**: Consider a specific time point $t_E \in \mathbb{T}_E$ and use the Lagrange multiplier $\pi$ to adjoin Eq. (7) with the performance index (5). Then $g_i|_{t_E}$ is a PEEC if and only if $\pi_i \geq 0$. In particular, it is a Pseudo-PEEC when $\pi_i = 0$. Also, $g_i|_{t_E}$ is a NEEC if and only if $\pi_i < 0$.

**Proof**: The proof uses the sensitivity analysis method suggested in Refs. [19] and [20]. In Problem 1, the EC at time point $t_E$ may be written as

$$g_i\big(\mathbf{y}(t_E), \mathbf{p}, t_E\big) = a_i, a_i = 0 \qquad (24)$$

Then the augmented performance index (only the relevant terms are listed) with the Langrage multiplier $\pi_i$ is

$$\bar{J}_g = J_g + \pi_i(g_i - a_i) + \ldots \qquad (25)$$

According to the sensitivity analysis method, Differentiating Eq. (25) with $a_i$ gives

$$\frac{\mathrm{d}\bar{\hat{J}}_g}{\mathrm{d}a_i} = \frac{\mathrm{d}\hat{J}_g}{\mathrm{d}a_i} = -\pi_i \qquad (26)$$

According to Definition 1, we may establish the statement. ∎

From Theorem 3, we may directly determine the type of an IEC from the multiplier information of its strengthened EC, without the need of substituting optimized solutions into the IEC for verification. In addition, for a Pseudo-PEEC, the corresponding Lagrange multiplier $\pi_i$ is $\pi_i = 0$. Consider the optimization problem without this Pseudo-PEEC, we may find the optimality conditions that the optimal solution satisfies are still same to the previous conditions. Thus we have

**Remark 4**: Omitting a Pseudo-PEEC from the optimization problem will not change the optimal solution.

Return to the IECs and consider an active IEC which may be equivalently written as

$$C_i\big(\mathbf{y}(t_I), \mathbf{p}, t_I\big) - a_i \leq 0, a_i = 0 \qquad (27)$$

Now the augmented performance index using the KKT multiplier $\pi_i$ is

$$\bar{J}_g = J_g + \pi_i(C_i - a_i) + \ldots \qquad (28)$$

Similarly with the sensitivity analysis method, Differentiating Eq. (28) with $a_i$ gives

$$\frac{\mathrm{d}\bar{\hat{J}}_g}{\mathrm{d}a_i} = \frac{\mathrm{d}\hat{J}_g}{\mathrm{d}a_i} = -\pi_i \qquad (29)$$

From the well-known property of KKT multiplier, we have $\pi_i \geq 0$, and this is consistent with the results in Theorem 3 for the strengthened PEEC.

## IV. VEM FOR OCPs INCLUDING TERMINAL IECs

*A. Problem definition*

In this paper, we consider the OCPs including terminal IECs, which are defined as

**Problem 2:** Consider performance index of Bolza form

$$J = \varphi(\bm{x}(t_f), t_f) + \int_{t_0}^{t_f} L(\bm{x}(t), \bm{u}(t), t) \mathrm{d}t \tag{30}$$

subject to the dynamic equation

$$\dot{\bm{x}} = \bm{f}(\bm{x}, \bm{u}, t) \tag{31}$$

where $t \in \mathbb{R}$ is the time. $\bm{x} \in \mathbb{R}^n$ is the state vector and its elements belong to $C^2[t_0, t_f]$. $\bm{u} \in \mathbb{R}^m$ is the control vector and its elements belong to $C^1[t_0, t_f]$. The function $L: \mathbb{R}^n \times \mathbb{R}^m \times \mathbb{R} \to \mathbb{R}$ and its first-order partial derivatives are continuous with respect to $\bm{x}$, $\bm{u}$ and $t$. The function $\varphi: \mathbb{R}^m \times \mathbb{R} \to \mathbb{R}$ and its first-order and second-order partial derivatives are continuous with respect to $\bm{x}$ and $t$. The vector function $\bm{f}: \mathbb{R}^n \times \mathbb{R}^m \times \mathbb{R} \to \mathbb{R}^n$ and its first-order partial derivatives are continuous and Lipschitz in $\bm{x}$, $\bm{u}$ and $t$. The initial time $t_0$ is fixed and the terminal time $t_f$ is free. The initial and terminal boundary conditions are respectively prescribed as

$$\bm{x}(t_0) = \bm{x}_0 \tag{32}$$

$$\bm{g}_E\left(\bm{x}(t_f), t_f\right) = \bm{0} \tag{33}$$

$$\bm{g}_I\left(\bm{x}(t_f), t_f\right) \leq \bm{0} \tag{34}$$

where $\bm{g}_E: \mathbb{R}^n \times \mathbb{R} \to \mathbb{R}^{q_E}$ is a $q_E$ dimensional vector function with continuous first-order partial derivatives, and $\bm{g}_I: \mathbb{R}^n \times \mathbb{R} \to \mathbb{R}^{q_I}$ is a $q_I$ dimensional vector function with continuous first-order partial derivatives. Find the optimal solution $(\hat{\bm{x}}, \hat{\bm{u}})$ that minimizes $J$, i.e.

$$(\hat{\bm{x}}, \hat{\bm{u}}) = \arg\min(J) \tag{35}$$

*B. Derivation of variation dynamic evolution equations*

In the way similar to Ref. [13], we consider the problem within the feasible solution domain $\mathbb{D}_o$, in which any solution satisfies Eqs.(31)-(34). The Bolza performance index (30) is first transformed to the equivalent Lagrange type as

$$J = \int_{t_0}^{t_f} \left(\varphi_t + \varphi_{\bm{x}}^{\mathrm{T}} \bm{f}(\bm{x}, \bm{u}, t) + L(\bm{x}, \bm{u}, t)\right) \mathrm{d}t \tag{36}$$

where $\varphi_t$ and $\varphi_{\bm{x}}$ are the partial derivatives. Differentiating Eq. (36) with respect to the variation time $\tau$ gives

$$\frac{\delta J}{\delta \tau} = (\varphi_t + \varphi_{\bm{x}}^{\mathrm{T}} \bm{f} + L)\Big|_{t_f} \frac{\delta t_f}{\delta \tau} + \int_{t_0}^{t_f} \left((\varphi_{t\bm{x}}^{\mathrm{T}} + \bm{f}^{\mathrm{T}} \varphi_{\bm{xx}} + \varphi_{\bm{x}}^{\mathrm{T}} \bm{f}_{\bm{x}} + L_{\bm{x}}^{\mathrm{T}}) \frac{\delta \bm{x}}{\delta \tau} + (\varphi_{\bm{x}}^{\mathrm{T}} \bm{f}_{\bm{u}} + L_{\bm{u}}^{\mathrm{T}}) \frac{\delta \bm{u}}{\delta \tau}\right) \mathrm{d}t \tag{37}$$

where $\varphi_{t\bm{x}}$ and $\varphi_{\bm{xx}}$ are second-order partial derivatives in the form of (column) vector and matrix, and $\bm{f}_{\bm{x}}$ and $\bm{f}_{\bm{u}}$ are the Jacobi matrixes. For the solutions in $\mathbb{D}_o$, $\frac{\delta \bm{x}}{\delta \tau}$ and $\frac{\delta \bm{u}}{\delta \tau}$ are related because of Eq. (31), and they need to satisfies the following variation equation as

$$\frac{\delta \dot{\bm{x}}}{\delta \tau} = \bm{f}_{\bm{x}} \frac{\delta \bm{x}}{\delta \tau} + \bm{f}_{\bm{u}} \frac{\delta \bm{u}}{\delta \tau} \tag{38}$$

with the initial condition $\left.\dfrac{\delta x}{\delta \tau}\right|_{t_0} = \mathbf{0}$. Note that $f_x$ and $f_u$ are time-dependent matrixes linearized at the feasible solution $x(t)$ and $u(t)$. Eq. (38) is a linear time-varying equation and has a zero initial value. Thus according to the linear system theory [21], its solution may be explicitly expressed as

$$\frac{\delta x}{\delta \tau} = \int_{t_0}^{t} \boldsymbol{\Phi}_o(t,s) f_u(s) \frac{\delta u}{\delta \tau}(s)\,\mathrm{d}s \tag{39}$$

where $\boldsymbol{\Phi}_o(t,s)$ is the $n \times n$ dimensional state transition matrix from time point $s$ to time point $t$. It satisfies

$$\frac{\partial}{\partial t}\boldsymbol{\Phi}_o(t,s) = f_x(t)\boldsymbol{\Phi}_o(t,s) \tag{40}$$

Substitute Eq. (39) into Eq. (37) and follow the same derivation as Ref. [13], we again obtain

$$\frac{\delta J}{\delta \tau} = \left.(\varphi_t + \varphi_x^{\mathrm{T}} f + L)\right|_{t_f} \frac{\delta t_f}{\delta \tau} + \int_{t_0}^{t_f} p_u^{\mathrm{T}} \frac{\delta u}{\delta \tau}\,\mathrm{d}t \tag{41}$$

where

$$p_u(t) = L_u + f_u^{\mathrm{T}} \varphi_x + f_u^{\mathrm{T}} \left( \int_t^{t_f} \boldsymbol{\Phi}_o^{\mathrm{T}}(\sigma,t)\left(L_x(\sigma) + \varphi_{tx}(\sigma) + \varphi_{xx}^{\mathrm{T}}(\sigma) f(\sigma) + f_x(\sigma)^{\mathrm{T}} \varphi_x(\sigma)\right)\mathrm{d}\sigma \right) \tag{42}$$

Now we will find the variation dynamic evolution equations that not only guarantees $\dfrac{\delta J}{\delta \tau} \le 0$ but also satisfies the variation equation for the terminal EC (33) as

$$\frac{\delta g_E}{\delta \tau} = (g_E)_{x_f} \int_{t_0}^{t_f} \boldsymbol{\Phi}_o(t_f,t) f_u(t) \frac{\delta u}{\delta \tau}(t)\,\mathrm{d}t + \left((g_E)_{x_f} f(t_f) + (g_E)_{t_f}\right)\frac{\delta t_f}{\delta \tau} = \mathbf{0} \tag{43}$$

and the variation motion allowed by the terminal IEC (34), i.e.

$$\frac{\delta (g_I)_i}{\delta \tau} = \left(\frac{\partial (g_I)_i}{\partial x_f}\right)^{\mathrm{T}} \int_{t_0}^{t_f} \boldsymbol{\Phi}_o(t_f,t) f_u(t) \frac{\delta u}{\delta \tau}(t)\,\mathrm{d}t + \left(\left(\frac{\partial (g_I)_i}{\partial x_f}\right)^{\mathrm{T}} f(t_f) + \frac{\partial (g_I)_i}{\partial t_f}\right)\frac{\delta t_f}{\delta \tau} \le 0 \quad i \in \mathbb{I} \tag{44}$$

where $\mathbb{I}$ is the index set of active IECs defined as

$$\mathbb{I} = \{i\,|\,(g_I)_i = 0,\ i = 1,2,...,q_I\} \tag{45}$$

In Ref. [14], we formulate a Multi-objective Optimization Problem (MOP) to derive the evolution equations that guarantees $\dfrac{\delta J}{\delta \tau} \le 0$ and Eq. (43), and to solve the Pareto optimal solution of this MOP, the weighting method is employed to get the Feasibility-preserving Evolution Optimization Problem (FPEOP) as

**Problem 3**:

$$\begin{aligned}
\min\quad & J_{t3} = \frac{1}{2}J_{t1} + \frac{1}{2}J_{t2} \\
\text{s.t.}\quad & \\
& \frac{\delta g_E}{\delta \tau} = \mathbf{0}
\end{aligned} \tag{46}$$

where

$$J_{t1} = \left.(\varphi_t + \varphi_x^{\mathrm{T}} f + L)\right|_{t_f} \frac{\delta t_f}{\delta \tau} + \int_{t_0}^{t_f} p_u^{\mathrm{T}} \frac{\delta u}{\delta \tau}\,\mathrm{d}t \tag{47}$$

$$J_{t2} = \frac{1}{2k_{t_f}}(\frac{\delta t_f}{\delta \tau})^2 + \int_{t_0}^{t_f} \frac{1}{2}(\frac{\delta \boldsymbol{u}}{\delta \tau})^T \boldsymbol{K}^{-1} \frac{\delta \boldsymbol{u}}{\delta \tau} \mathrm{d}t \tag{48}$$

with $\frac{\delta \boldsymbol{u}}{\delta \tau}$ being the optimization variable and $\frac{\delta t_f}{\delta \tau}$ being the optimization parameter. $\boldsymbol{K}$ is a $m \times m$ dimensional positive-definite matrix and $k_{t_f}$ is a positive constant.

Due to the variation constraint (44) arising from the terminal IECs (34), extra considerations are required on constructing the FPEOP. However, we cannot simply set that $\frac{\delta(g_I)_i}{\delta \tau} = 0$ for all $i \in \mathbb{I}$ to form the optimization problem, because such treatment may produce the wrong solution. In order to obtain the right evolution equations that may seek the optimal solution, only the active IECs in (44) need to be considered. For the IECs that are inactive, they will return to the inactive domain automatically. Thus we construct the following FPEOP as

**Problem 4**:

$$\begin{aligned} \min \quad & J_{t3} = \frac{1}{2}J_{t1} + \frac{1}{2}J_{t2} \\ s.t. \quad & \\ & \frac{\delta \boldsymbol{g}_E}{\delta \tau} = \boldsymbol{0} \\ & \frac{\delta (g_I)_i}{\delta \tau} = 0 \quad i \in \mathbb{I}_p \end{aligned} \tag{49}$$

where the index set $\mathbb{I}_p$ is a subset of $\mathbb{I}$ defined as

$$\mathbb{I}_p = \{i \,|\, (g_I)_i = 0, \frac{\delta(g_I)_i}{\delta \tau} \leq 0 \text{ is an active IEC, } i = 1, 2, ..., q_I\} \tag{50}$$

with the number of its elements denoted by $n_{\mathbb{I}_p}$.

In the way same to Ref. [14], through solving Problem 4 analytically, we may obtain the variation dynamic evolution equations for Problem 2. See the following theorem.

**Theorem 4**: The following variation dynamic evolution equations guarantees that the solution stays in the feasible domain and the performance index $\frac{\delta J}{\delta \tau} \leq 0$

$$\frac{\delta \boldsymbol{u}}{\delta \tau} = -\boldsymbol{K}\left(\boldsymbol{p_u} + \boldsymbol{f_u}^T \boldsymbol{\Phi}^T(t_f, t)\left\{(\boldsymbol{g}_E)_{\boldsymbol{x}_f}^T \boldsymbol{\pi}_E + (\boldsymbol{g}_I)_{\boldsymbol{x}_f}^T \boldsymbol{\pi}_I\right\}\right) \tag{51}$$

$$\frac{\delta t_f}{\delta \tau} = -k_{t_f}\left(\varphi_t + \varphi_{\boldsymbol{x}}^T \boldsymbol{f} + L + \boldsymbol{\pi}_E^T\left((\boldsymbol{g}_E)_{\boldsymbol{x}_f} \boldsymbol{f} + (\boldsymbol{g}_E)_{t_f}\right) + \boldsymbol{\pi}_I^T\left((\boldsymbol{g}_I)_{\boldsymbol{x}_f} \boldsymbol{f} + (\boldsymbol{g}_I)_{t_f}\right)\right)\bigg|_{t_f} \tag{52}$$

where $\boldsymbol{K}$ is the $m \times m$ dimensional positive-definite matrix and $k_{t_f}$ is a positive constant, $\boldsymbol{p_u}$ is defined in Eq. (42) and the parameter vectors $\boldsymbol{\pi}_E \in \mathbb{R}^{q_E}$ and $\boldsymbol{\pi}_I \in \mathbb{R}^{q_I}$ are determined by

$$\pi_E = \begin{bmatrix} \pi_1 \\ \pi_2 \\ \cdots \\ \pi_{q_E} \end{bmatrix} \tag{53}$$

$$(\pi_I)_i = 0 \quad i \notin \mathbb{I}_p, \quad \pi_I(\mathbb{I}_p) = \begin{bmatrix} \pi_{q_E+1} \\ \pi_{q_E+2} \\ \cdots \\ \pi_{q_E+n_{\mathbb{I}_p}} \end{bmatrix} \geq \mathbf{0}$$

where the parameter $\boldsymbol{\pi} \in \mathbb{R}^{q_E+n_{\mathbb{I}_p}}$ is the solution of the linear matrix equation

$$\boldsymbol{M\pi} = -\boldsymbol{r} \tag{54}$$

The $(q_E+n_{\mathbb{I}_p}) \times (q_E+n_{\mathbb{I}_p})$ dimensional matrix $\boldsymbol{M}$ and the $q_E+n_{\mathbb{I}_p}$ dimensional vector $\boldsymbol{r}$ are

$$\boldsymbol{M} = \boldsymbol{g}_{\boldsymbol{x}_f} \left( \int_{t_0}^{t_f} \boldsymbol{\Phi}(t_f,t) \boldsymbol{f_u} \boldsymbol{K} \boldsymbol{f_u}^{\mathrm{T}} \boldsymbol{\Phi}^{\mathrm{T}}(t_f,t) \mathrm{d}t \right) \boldsymbol{g}_{\boldsymbol{x}_f}^{\mathrm{T}} + k_{t_f}(\boldsymbol{g}_{\boldsymbol{x}_f} \boldsymbol{f} + \boldsymbol{g}_{t_f})(\boldsymbol{g}_{\boldsymbol{x}_f} \boldsymbol{f} + \boldsymbol{g}_{t_f})^{\mathrm{T}} \Big|_{t_f} \tag{55}$$

$$\boldsymbol{r} = \boldsymbol{g}_{\boldsymbol{x}_f} \left( \int_{t_0}^{t_f} \boldsymbol{\Phi}(t_f,t) \boldsymbol{f_u} \boldsymbol{K} \boldsymbol{p_u} \mathrm{d}t \right) + k_{t_f}(\boldsymbol{g}_{\boldsymbol{x}_f} \boldsymbol{f} + \boldsymbol{g}_{t_f})(\varphi_t + \varphi_{\boldsymbol{x}}^{\mathrm{T}} \boldsymbol{f} + L) \Big|_{t_f} \tag{56}$$

with $\boldsymbol{g} = \begin{bmatrix} \boldsymbol{g}_E \\ \boldsymbol{g}_I(\mathbb{I}_p) \end{bmatrix}$. Moreover, under the evolution equations (51) and (52), $\dfrac{\delta J}{\delta \tau} = 0$ occurs only when

$$\boldsymbol{p_u} + \boldsymbol{f_u}^{\mathrm{T}} \boldsymbol{\Phi}^{\mathrm{T}}(t_f,t)\left( (\boldsymbol{g}_E)_{\boldsymbol{x}_f}^{\mathrm{T}} \boldsymbol{\pi}_E + (\boldsymbol{g}_I)_{\boldsymbol{x}_f}^{\mathrm{T}} \boldsymbol{\pi}_I \right) = \boldsymbol{0} \tag{57}$$

$$\varphi_t + \varphi_{\boldsymbol{x}}^{\mathrm{T}} \boldsymbol{f} + L + \boldsymbol{\pi}_E^{\mathrm{T}}\left( (\boldsymbol{g}_E)_{\boldsymbol{x}_f} \boldsymbol{f} + (\boldsymbol{g}_E)_{t_f} \right) + \boldsymbol{\pi}_I^{\mathrm{T}}\left( (\boldsymbol{g}_I)_{\boldsymbol{x}_f} \boldsymbol{f} + (\boldsymbol{g}_I)_{t_f} \right) = 0 \tag{58}$$

For the optimal solution, there is $\mathbb{I}_p = \mathbb{I}$, and the optimal value of $\boldsymbol{\pi}$ (corresponding to the right $\mathbb{I}_p$) satisfies

$$\begin{bmatrix} \boldsymbol{M}_{s1} \\ \boldsymbol{M}_{s2} \end{bmatrix} \boldsymbol{\pi} = -\begin{bmatrix} \boldsymbol{r}_{s1} \\ \boldsymbol{r}_{s2} \end{bmatrix} \tag{59}$$

where the $(q_E+n_{\mathbb{I}_p}) \times (q_E+n_{\mathbb{I}_p})$ dimensional matrixes $\boldsymbol{M}_{s1}$, $\boldsymbol{M}_{s2}$ and the $q_E+n_{\mathbb{I}_p}$ dimensional vectors $\boldsymbol{r}_{s1}$, $\boldsymbol{r}_{s2}$ are

$$\boldsymbol{M}_{s1} = \boldsymbol{g}_{\boldsymbol{x}_f} \left( \int_{t_0}^{t_f} \boldsymbol{\Phi}(t_f,t) \boldsymbol{f_u} \boldsymbol{f_u}^{\mathrm{T}} \boldsymbol{\Phi}^{\mathrm{T}}(t_f,t) \mathrm{d}t \right) \boldsymbol{g}_{\boldsymbol{x}_f}^{\mathrm{T}} \tag{60}$$

$$\boldsymbol{M}_{s2} = (\boldsymbol{g}_{\boldsymbol{x}_f} \boldsymbol{f} + \boldsymbol{g}_{t_f})(\boldsymbol{g}_{\boldsymbol{x}_f} \boldsymbol{f} + \boldsymbol{g}_{t_f})^{\mathrm{T}} \Big|_{t_f} \tag{61}$$

$$\boldsymbol{r}_{s1} = \boldsymbol{g}_{\boldsymbol{x}_f} \int_{t_0}^{t_f} \boldsymbol{\Phi}(t_f,t) \boldsymbol{f_u} \boldsymbol{p_u} \mathrm{d}t \tag{62}$$

$$\boldsymbol{r}_{s2} = (\boldsymbol{g}_{\boldsymbol{x}_f} \boldsymbol{f} + \boldsymbol{g}_{t_f})(\varphi_t + \varphi_{\boldsymbol{x}}^{\mathrm{T}} \boldsymbol{f} + L) \Big|_{t_f} \tag{63}$$

The proof of Theorem 4 is similar to that of Ref. [14]. Regarding the argument that $\mathbb{I}_p = \mathbb{I}$ for the optimal solution of Problem 2, this is because any component in $\mathbb{I}$ also belongs to $\mathbb{I}_p$ ultimately, or this active IEC will become inactive. During the evolution process, the set $\mathbb{I}_p$ needs to be determined. Generally $\mathbb{I}$ is easy to get. Thus we may first strengthen all IECs in $\mathbb{I}$ to get the corresponding Lagrange multipliers, and then use Theorem 3 to select the right $\mathbb{I}_p$. Also, for the linear equation (54), assuming that the control satisfies the controllability requirement [22], then the solution is guaranteed. When $M$ is invertible, the parameter $\pi$ may be calculated as

$$\pi = -M^{-1}r \tag{64}$$

*C. Equivalence to the classic optimality conditions*

Actually, Eqs. (57) and (58) are the first-order costate-free optimality conditions for Problem 2. We will show that they are equivalent to the traditional ones with costates [23]. By the adjoining method, we may constructed the augmented functional as

$$\bar{J} = \varphi(x(t_f), t_f) + \bar{\pi}_E^{\mathrm{T}} g_E\left(x(t_f), t\right) + \bar{\pi}_I^{\mathrm{T}} g_I\left(x(t_f), t\right) + \int_{t_0}^{t_f} \left(L + \lambda^{\mathrm{T}}(f - \dot{x})\right) \mathrm{d}t \tag{65}$$

where $\lambda \in \mathbb{R}^n$ is the costate variable vector, $\bar{\pi}_E \in \mathbb{R}^{q_E}$ is Lagrange multiplier parameter vector, and $\bar{\pi}_I \in \mathbb{R}^{q_I}$ is KKT multiplier parameter vector. Then the corresponding first-order variation may be derived as

$$\delta \bar{J} = \left(\varphi_{t_f} + \bar{\pi}_E^{\mathrm{T}}(g_E)_{t_f} + \bar{\pi}_I^{\mathrm{T}}(g_I)_{t_f} + H\right)\bigg|_{t_f} \delta t_f + \left(\lambda(t_f) - \varphi_x(t_f) - (g_E)_{x_f}^{\mathrm{T}} \bar{\pi}_E - (g_I)_{x_f}^{\mathrm{T}} \bar{\pi}_I\right) \delta x(t_f)$$
$$+ \int_{t_0}^{t_f} \left((H_\lambda - \dot{x})^{\mathrm{T}} \delta \lambda + (H_x + \dot{\lambda})^{\mathrm{T}} \delta x + H_u^{\mathrm{T}} \delta u\right) \mathrm{d}t \tag{66}$$

with

$$\begin{aligned} (\bar{\pi}_I)_i (g_I)_i &= 0 \quad & i \in \mathbb{I} \\ (\bar{\pi}_I)_i &= 0 \quad & i \notin \mathbb{I} \end{aligned} \tag{67}$$

where $H = L + \lambda^{\mathrm{T}} f$ is the Hamiltonian. Through $\delta \bar{J} = 0$, we have

$$\dot{\lambda} + H_x = \dot{\lambda} + L_x + f_x^{\mathrm{T}} \lambda = 0 \tag{68}$$

$$H_u = L_u + f_u^{\mathrm{T}} \lambda = 0 \tag{69}$$

and the transversality conditions

$$H(t_f) + \varphi_{t_f} + \bar{\pi}_E^{\mathrm{T}}(g_E)_{t_f} + \bar{\pi}_I^{\mathrm{T}}(g_I)_{t_f} = 0 \tag{70}$$

$$\lambda(t_f) - \varphi_x(t_f) - (g_E)_{x_f}^{\mathrm{T}} \bar{\pi}_E - (g_I)_{x_f}^{\mathrm{T}} \bar{\pi}_I = 0 \tag{71}$$

**Theorem 5:** For Problem 2, the optimality conditions given by Eqs. (57) and (58) are equivalent to the optimality conditions given by (68)-(71).

**Proof**: Define a quantity $\gamma(t)$ as

$$\gamma(t) = \varphi_x(t) + \Phi_o^{\mathrm{T}}(t_f, t)\left((g_E)_{x_f}^{\mathrm{T}} \pi_E + (g_I)_{x_f}^{\mathrm{T}} \pi_I\right) + \int_t^{t_f} \Phi_o^{\mathrm{T}}(\sigma, t)\left(L_x(\sigma) + \varphi_{tx}(\sigma) + \varphi_{xx}^{\mathrm{T}}(\sigma) f(\sigma) + f_x(\sigma)^{\mathrm{T}} \varphi_x(\sigma)\right) \mathrm{d}\sigma \tag{72}$$

Then Eq. (57) is simplified as

$$L_u + f_u^{\mathrm{T}} \gamma = 0 \tag{73}$$

Obviously, when $t = t_f$, there is

$$\gamma(t_f) = \varphi_x(t_f) + (g_E)_{x_f}^{\mathrm{T}} \pi_E + (g_I)_{x_f}^{\mathrm{T}} \pi_I \tag{74}$$

Differentiate $\gamma(t)$ with respect to $t$. In the process, we will use the Leibniz rule [24]

$$\frac{\mathrm{d}}{\mathrm{d}t}\left(\int_{b(t)}^{a(t)} h(\sigma, t) \mathrm{d}\sigma\right) = h(a(t), t)\frac{\mathrm{d}}{\mathrm{d}t} a(t) - h(b(t), t)\frac{\mathrm{d}}{\mathrm{d}t} b(t) + \int_{b(t)}^{a(t)} h_t(\sigma, t) \mathrm{d}\sigma \tag{75}$$

and the property of $\boldsymbol{\Phi}_o(\sigma,t)$ [21]

$$\frac{\partial \boldsymbol{\Phi}_o(\sigma,t)}{\partial t} = -\boldsymbol{\Phi}_o(\sigma,t)\boldsymbol{f}_x(t) \quad (76)$$

$$\boldsymbol{\Phi}_o(t,t) = \boldsymbol{1} \quad (77)$$

where $\boldsymbol{1}$ is the $n \times n$ dimensional identity matrix. Then we have

$$\begin{aligned}\frac{d}{dt}\gamma(t) &= \varphi_{tx} + \varphi_{xx}^T \boldsymbol{f} - \boldsymbol{f}_x^T \boldsymbol{\Phi}_o^T(t_f,t)\left((\boldsymbol{g}_E)_{x_f}^T \boldsymbol{\pi}_E + (\boldsymbol{g}_I)_{x_f}^T \boldsymbol{\pi}_I\right) - \left(L_x + \varphi_{tx} + \varphi_{xx}^T \boldsymbol{f} + \boldsymbol{f}_x^T \varphi_x\right) - \boldsymbol{f}_x^T \int_t^{t_f} \boldsymbol{\Phi}_o^T(\sigma,t)\left(L_x(\sigma) + \varphi_{tx}(\sigma) + \varphi_{xx}^T(\sigma)\boldsymbol{f}(\sigma) + \boldsymbol{f}_x(\sigma)^T \varphi_x(\sigma)\right)d\sigma \\ &= -L_x - \boldsymbol{f}_x^T\left(\varphi_x(t) + \boldsymbol{\Phi}_o^T(t_f,t)\left((\boldsymbol{g}_E)_{x_f}^T \boldsymbol{\pi}_E + (\boldsymbol{g}_I)_{x_f}^T \boldsymbol{\pi}_I\right) + \int_t^{t_f} \boldsymbol{\Phi}_o^T(\sigma,t)\left(L_x(\sigma) + \varphi_{tx}(\sigma) + \varphi_{xx}^T(\sigma)\boldsymbol{f}(\sigma) + \boldsymbol{f}_x(\sigma)^T \varphi_x(\sigma)\right)d\sigma\right) \\ &= -L_x - \boldsymbol{f}_x^T \gamma(t)\end{aligned} \quad (78)$$

This means $\gamma(t)$ conforms to the same dynamics as the costates $\lambda(t)$ in Eq. (68). Furthermore, Eq. (41) may be reformulated as

$$\begin{aligned}\frac{\delta J}{\delta \tau} &= \left(L + \varphi_t + \varphi_x^T \boldsymbol{f} + \boldsymbol{\pi}_E^T\left((\boldsymbol{g}_E)_{x_f}\boldsymbol{f} + (\boldsymbol{g}_E)_{t_f}\right) + \boldsymbol{\pi}_I^T\left((\boldsymbol{g}_I)_{x_f}\boldsymbol{f} + (\boldsymbol{g}_I)_{t_f}\right)\right)\bigg|_{t_f} \frac{\delta t_f}{\delta \tau} \\ &\quad + \int_{t_0}^{t_f}\left(\boldsymbol{p}_u + \boldsymbol{f}_u^T \boldsymbol{\Phi}^T(t_f,t)\left\{(\boldsymbol{g}_E)_{x_f}^T \boldsymbol{\pi}_E + (\boldsymbol{g}_I)_{x_f}^T \boldsymbol{\pi}_I\right\}\right)^T \frac{\delta \boldsymbol{u}}{\delta \tau} dt \\ &\quad - \boldsymbol{\pi}_E^T\left((\boldsymbol{g}_E)_{x_f}\int_{t_0}^{t_f}\boldsymbol{\Phi}(t_f,t)\boldsymbol{f}_u\frac{\delta \boldsymbol{u}}{\delta \tau}dt + \left((\boldsymbol{g}_E)_{x_f}\boldsymbol{f} + (\boldsymbol{g}_E)_{t_f}\right)\bigg|_{t_f}\frac{\delta t_f}{\delta \tau}\right) \\ &\quad - \boldsymbol{\pi}_I^T\left((\boldsymbol{g}_I)_{x_f}\int_{t_0}^{t_f}\boldsymbol{\Phi}(t_f,t)\boldsymbol{f}_u\frac{\delta \boldsymbol{u}}{\delta \tau}dt + \left((\boldsymbol{g}_I)_{x_f}\boldsymbol{f} + (\boldsymbol{g}_I)_{t_f}\right)\bigg|_{t_f}\frac{\delta t_f}{\delta \tau}\right)\end{aligned} \quad (79)$$

Since under Eqs. (51) and (52), the last two terms in the right part of Eq. (79) vanish. Then

$$\frac{\delta J}{\delta \tau} = \left(L + \varphi_t + \varphi_x^T \boldsymbol{f} + \boldsymbol{\pi}_E^T\left((\boldsymbol{g}_E)_{x_f}\boldsymbol{f} + (\boldsymbol{g}_E)_{t_f}\right) + \boldsymbol{\pi}_I^T\left((\boldsymbol{g}_I)_{x_f}\boldsymbol{f} + (\boldsymbol{g}_I)_{t_f}\right)\right)\bigg|_{t_f}\frac{\delta t_f}{\delta \tau} + \int_{t_0}^{t_f}\left(\boldsymbol{p}_u + \boldsymbol{f}_u^T \boldsymbol{\Phi}^T(t_f,t)\left\{(\boldsymbol{g}_E)_{x_f}^T \boldsymbol{\pi}_E + (\boldsymbol{g}_I)_{x_f}^T \boldsymbol{\pi}_I\right\}\right)^T\frac{\delta \boldsymbol{u}}{\delta \tau}dt \quad (80)$$

which hold in the feasible solution domain $\mathbb{D}_o$. Further combined with Eq. (74) and ignore $\delta \tau$, we have

$$\delta J = \left(L + \varphi_t + \boldsymbol{\pi}_E^T(\boldsymbol{g}_E)_{t_f} + \boldsymbol{\pi}_I^T(\boldsymbol{g}_I)_{t_f} + \gamma(t_f)^T \boldsymbol{f}\right)\bigg|_{t_f}\delta t_f + \int_{t_0}^{t_f}\left(\boldsymbol{p}_u + \boldsymbol{f}_u^T \boldsymbol{\Phi}^T(t_f,t)\left\{(\boldsymbol{g}_E)_{x_f}^T \boldsymbol{\pi}_E + (\boldsymbol{g}_I)_{x_f}^T \boldsymbol{\pi}_I\right\}\right)^T\delta \boldsymbol{u}\,dt \quad (81)$$

Eq. (81) obviously hold at the optimal solution. Compare Eq. (66) with Eq. (81), because $\delta t_f$ may be arbitrary small, to achieve the extremal condition, Eqs. (58) and (70) should be same, i.e.

$$\left(L + \varphi_{t_f} + \bar{\boldsymbol{\pi}}_E^T(\boldsymbol{g}_E)_{t_f} + \bar{\boldsymbol{\pi}}_I^T(\boldsymbol{g}_I)_{t_f} + \boldsymbol{\lambda}^T \boldsymbol{f}\right)\bigg|_{t_f} = \left(L + \varphi_t + \boldsymbol{\pi}_E^T(\boldsymbol{g}_E)_{t_f} + \boldsymbol{\pi}_I^T(\boldsymbol{g}_I)_{t_f} + \gamma(t_f)^T \boldsymbol{f}\right)\bigg|_{t_f} \quad (82)$$

Since Eq. (82) generally hold for arbitrary $\boldsymbol{g}_E$, $\boldsymbol{g}_I$ and $\boldsymbol{f}$, and $\mathbb{I}_p = \mathbb{I}$ for the optimal solution, we can conclude that

$$\boldsymbol{\pi}_E = \bar{\boldsymbol{\pi}}_E \quad (83)$$

$$\boldsymbol{\pi}_I = \bar{\boldsymbol{\pi}}_I \quad (84)$$

$$\gamma(t_f) = \lambda(t_f) \quad (85)$$

Therefore Eq.(74) is same to Eq. (71). With Eqs. (78) and (85), the relation that $\gamma(t) = \lambda(t)$ is established. Then Eq. (57) and Eq. (69) are identical. ∎

From Theorem 5, we get the explicit analytic relations between the costates $\lambda$, the multipliers $\bar{\boldsymbol{\pi}}_E$, $\bar{\boldsymbol{\pi}}_I$ in the classic treatment in Eq. (65) and the original (state and control) variables, which formerly can only be obtained numerically by solving the BVP. After the proof of Theorem 5, now the variables evolving direction using the VEM is easy to determine and the optimal solution of Problem 2 will be sought with theoretical guarantee.

**Theorem 6:** Solving the IVP with respect to $\tau$, defined by the variation dynamic evolution equations (39), (51) and (52) from a feasible initial solution, when $\tau \to +\infty$, $(x, u)$ will satisfy the optimality conditions of Problem 2.

Proof: By Lemma 1 and with Eq. (30) as the Lyapunov functional, we may claim that the minimum solution of Problem 2 is an asymptotically stable solution within the feasibility domain $\mathbb{D}_o$ for the infinite-dimensional dynamics governed by Eqs. (39), (51) and (52). From a feasible initial solution, any evolution under these dynamics maintains the feasibility of the variables, and they also guarantee $\frac{\delta J}{\delta \tau} \leq 0$. The functional $J$ will decrease until $\frac{\delta J}{\delta \tau} = 0$, which occurs when $\tau \to +\infty$ due to the asymptotical approach. When $\frac{\delta J}{\delta \tau} = 0$, this determines the optimal conditions, namely, Eqs. (57) and (58). ∎

*D. Formulation of EPDE*

Use the partial differential operator "$\partial$" and the differential operator "d" to reformulate the variation dynamic evolution equations (39), (51) and (52), we may get the EPDE and EDE as

$$\frac{\partial}{\partial \tau}\begin{bmatrix} x \\ u \end{bmatrix} = \begin{bmatrix} \int_{t_0}^{t} \boldsymbol{\Phi}_o(t,s) \boldsymbol{f}_u(s) \frac{\partial \boldsymbol{u}}{\partial \tau}(s) \, \mathrm{d}s \\ -\boldsymbol{K}\left(\boldsymbol{p}_u + \boldsymbol{f}_u^{\mathrm{T}} \boldsymbol{\Phi}^{\mathrm{T}}(t_f, t)\left\{(\boldsymbol{g}_E)_{x_f}^{\mathrm{T}} \boldsymbol{\pi}_E + (\boldsymbol{g}_I)_{x_f}^{\mathrm{T}} \boldsymbol{\pi}_I\right\}\right) \end{bmatrix} \tag{86}$$

$$\frac{\mathrm{d}t_f}{\mathrm{d}\tau} = -k_{t_f}\left(L + \varphi_t + \varphi_x^{\mathrm{T}} \boldsymbol{f} + \boldsymbol{\pi}_E^{\mathrm{T}}\left((\boldsymbol{g}_E)_{x_f} \boldsymbol{f} + (\boldsymbol{g}_E)_{t_f}\right) + \boldsymbol{\pi}_I^{\mathrm{T}}\left((\boldsymbol{g}_I)_{x_f} \boldsymbol{f} + (\boldsymbol{g}_I)_{t_f}\right)\right)\bigg|_{t_f} \tag{87}$$

The definite conditions are $t_f|_{\tau=0} = \tilde{t}_f$ and $\begin{bmatrix} x(t,\tau) \\ u(t,\tau) \end{bmatrix}\bigg|_{\tau=0} = \begin{bmatrix} \tilde{x}(t) \\ \tilde{u}(t) \end{bmatrix}$, where $\tilde{x}(t)$ and $\tilde{u}(t)$ are the initial feasible solution.

Eqs. (86) and (87) realize the anticipated variable evolving along the variation time $\tau$ as depicted in Fig. 1. The initial conditions of $x(t,\tau)$ and $u(t,\tau)$ at $\tau = 0$ belong to the feasible solution domain and their value at $\tau = +\infty$ represents the optimal solution of the OCP. The right part of the EPDE (86) is also only a vector function of time $t$. Thus we may apply the semi-discrete method to discretize it along the normal time dimension and further use ODE integration methods to get the numerical solution. Moreover, the results obtained in this paper are also applicable to the OCPs with fixed terminal time. By setting $k_{t_f} = 0$, these equations may be directly applied.

*E. Numerical soft barrier*

Theoretically, the evolution equations will precisely seek the optimal solution. During the variable evolution process, once the IECs are activated, corresponding variation constraints will be triggered immediately to maintain the feasibility of solutions. However, since we resort to the numerical method for the solution, concretely by using the ODE integration methods to solve the transformed IVPs, the numerical error is unavoidable, and this may leads to the violation of the IECs. Refer to the strategy to turn an infeasible solution to be feasible in Ref. [25], a numerical soft barrier is introduced to remove the possible numerical error, by adapting the FPEOP as
**Problem 5**:

$$\min \quad J_{t3} = \frac{1}{2}J_{t1} + \frac{1}{2}J_{t2}$$
$$\text{s.t.}$$
$$\frac{\delta \boldsymbol{g}_E}{\delta \tau} = \boldsymbol{0} \tag{88}$$
$$\frac{\delta (g_I)_i}{\delta \tau} + k_{g_I}(g_I)_i = 0 \quad i \in \mathbb{I}_p$$

where $k_{g_I}$ is a positive constant and now the index set $\mathbb{I}_p$ is defined as

$$\mathbb{I}_p = \{i \,|\, (g_I)_i \geq 0, \frac{\delta (g_I)_i}{\delta \tau} + k_{g_I}(g_I)_i \leq 0 \text{ is an active IEC}, \ i = 1, 2, ..., q_I\} \tag{89}$$

Through solving Problem 5, the evolution equations derived are similar except Eq. (56) are modified as

$$\boldsymbol{r} = \boldsymbol{g}_{\boldsymbol{x}_f}\left(\int_{t_0}^{t_f} \boldsymbol{\Phi}(t_f, t)\boldsymbol{f_u}\boldsymbol{Kp_u}\,\mathrm{d}t\right) + k_{t_f}(\boldsymbol{g}_{\boldsymbol{x}_f}\boldsymbol{f} + \boldsymbol{g}_{t_f})(\varphi_t + \varphi_{\boldsymbol{x}}^{\mathrm{T}}\boldsymbol{f} + L)\Big|_{t_f} + k_{g_I}\begin{bmatrix} \boldsymbol{0} \\ \boldsymbol{g}_I(\mathbb{I}_P) \end{bmatrix} \tag{90}$$

In this way, the possible violations on the IECs due to the numerical error will be eliminated gradually.

## V. ILLUSTRATIVE EXAMPLE

We consider a nonlinear example adapted from the Brachistochrone problem [26], which describes the motion curve of the fastest descending. The dynamic equations are

$$\dot{\boldsymbol{x}} = \boldsymbol{f}(\boldsymbol{x}, u)$$

where $\boldsymbol{x} = \begin{bmatrix} x \\ y \\ V \end{bmatrix}$, $\boldsymbol{f} = \begin{bmatrix} V\sin(u) \\ -V\cos(u) \\ g\cos(u) \end{bmatrix}$, $g = 10$ is the gravity constant. Find the solution that minimizes the performance index

$$J = t_f$$

with the boundary conditions

$$\begin{bmatrix} x \\ y \\ V \end{bmatrix}\bigg|_{t_0 = 0} = \begin{bmatrix} 0 \\ 0 \\ 0 \end{bmatrix}, \quad \begin{matrix} x(t_f) = 2 \\ y(t_f) \leq -2 \end{matrix}$$

In the specific form of the EPDE (86) and the EDE (87), the parameters $K$ and $k_{t_f}$ were set to be 0.1 and 0.05, respectively. The barrier parameter $k_{g_I}$ in Eq. (90) was set to be 0.1. The definite conditions, i.e., $\begin{bmatrix} \boldsymbol{x}(t,\tau) \\ u(t,\tau) \\ t_f(\tau) \end{bmatrix}\bigg|_{\tau=0}$, were obtained from a physical motion along a straight line that connects the initial position to the terminal position of $\begin{bmatrix} 2 \\ -2\sqrt{3} \\ 0 \end{bmatrix}$, i. e.

$$\tilde{t}_f = \sqrt{\frac{8\sqrt{3}}{15}} \qquad \tilde{u} = \frac{\pi}{6}$$
$$\tilde{x} = \frac{5\sqrt{3}}{4}t^2 \qquad \tilde{y} = -\frac{15}{4}t^2 \qquad \tilde{V} = 5\sqrt{3}t$$

We discretized the time horizon $[t_0, t_f]$ uniformly, with 101 points. Thus, a large IVP with 405 states (including the terminal time) is obtained. We employed "ode45" in Matlab for the numerical integration. In the integrator setting, the default relative error tolerance and the absolute error tolerance are $1\times 10^{-3}$ and $1\times 10^{-6}$, respectively. For comparison, we computed the optimal solution with GPOPS-II [27], a Radau PS method based OCP solver.

Fig. 2 gives the states curve in the $xy$ coordinate plane, showing that the numerical results starting from a straight line approach the optimal solution over time. The control solutions are plotted in Fig. 3, and the asymptotical approach of the numerical results are demonstrated. In Fig. 4, the terminal time profile against the variation time $\tau$ is plotted. The result of $t_f$ declines rapidly at first and then gradually approaches to the minimum decline time, and it only changes slightly after $\tau = 40$s. At $\tau = 300$s, we compute that $t_f = 0.8165$s from the VEM, same to the result from GPOPS-II. Fig. 5 presents the evolution profiles of the Lagrange multiplier $\pi_E$ and the KKT multiplier $\pi_I$. It is shown that $\pi_I$ suddenly jumps to the value about 0.05 at $\tau = 8.9$s. At $\tau = 300$s, we have $\pi_E = -0.1477$ and $\pi_I = 0.0564$. In Fig 6, the profile of the terminal IEC on $y(t_f)$ is presented. The optimal solution finally approaches the upper limit and the IEC is active. Different from the curve of $\pi_I$, the change of $y(t_f)$ is continuous (although sharp) and the IEC is activated at $\tau = 8.9$s. In particular, from the close-up, it is noted that the IEC is violated due to the integration numerical error, and it gradually converges to the allowed value of -2, under the effect from the numerical soft barrier.

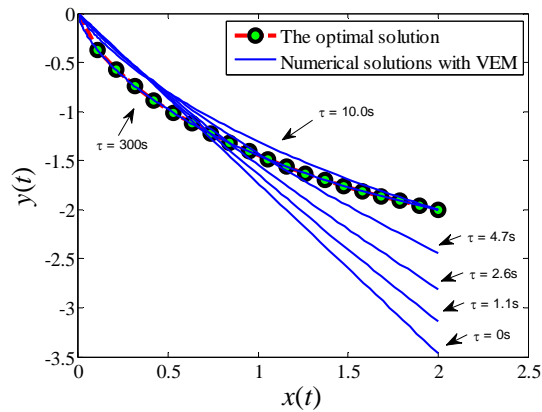

Fig. 2 The evolution of numerical solutions in the $xy$ coordinate plane to the optimal solution.

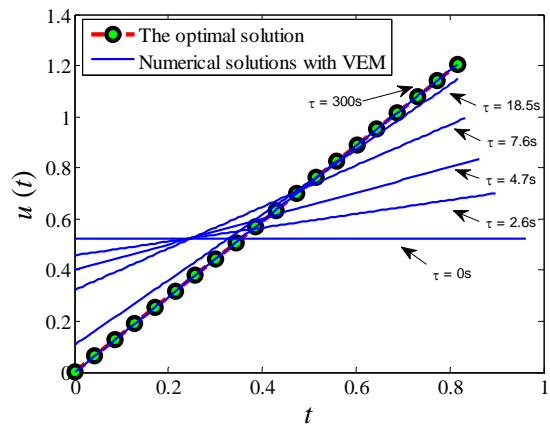

Fig. 3. The evolution of numerical solutions of $u$ to the optimal solution.

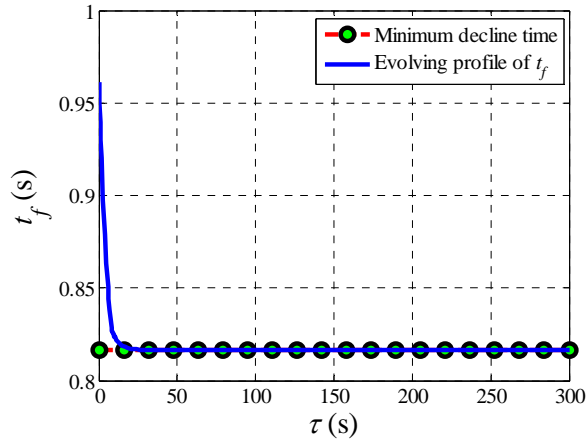

Fig. 4 The evolution profile of $t_f$ to the minimum decline time.

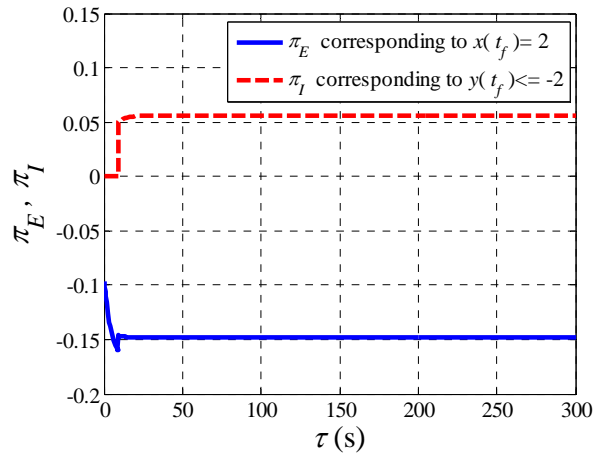

Fig. 5 The evolution profiles of Lagrange multiplier $\pi_E$ and KKT multiplier $\pi_I$.

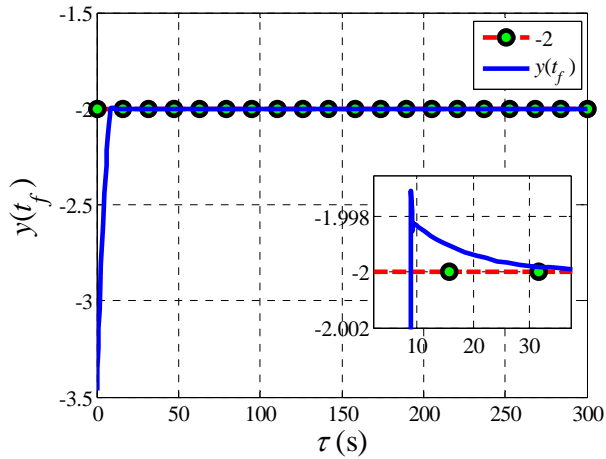

Fig. 6 The evolution profile of terminal IEC on $y(t_f)$.

In addition, we consider another version of this example, in which the terminal boundary conditions are specified as

$$x(t_f) = 2$$
$$-1.3 \leq y(t_f) \leq -1$$

To inspect the evolution detail of getting rid of activated IECs, the definite conditions are intentionally set as

$$\tilde{t}_f = 1 \qquad \tilde{u} = \arctan(2)$$
$$\tilde{x} = 2t^2 \qquad \tilde{y} = -t^2 \qquad \tilde{V} = 2\sqrt{5}t$$

which represents a physical motion along a straight line that connects the initial position to the terminal position of $\begin{bmatrix} 2 \\ -1 \\ 0 \end{bmatrix}$, while all the other settings are still the same. Fig. 7 gives the evolution profile of $y(t_f)$, showing that it is well constrained within the IEC bounds, and the activated IECs during the evolution ($y(t_f)+1 \leq 0$ at $\tau = 0$s and $-y(t_f)-1.3 \leq 0$ when $\tau \in [9.0\text{s}, 29.0\text{s}]$ specifically) are successfully got rid of. Combined with the information of KKT multipliers given in Fig. 8, it is found that the even if the IEC $y(t_f)+1 \leq 0$ is active at $\tau = 0$s, its variation equation (i.e., Eq. (44)) is an inactive IEC for the FPEOP, and the corresponding KKT multiplier is always zero. Regarding the multiplier for the IEC of $-y(t_f)-1.3 \leq 0$, its value suddenly jumped to about 0.01 at $\tau = 9.0$s. Then it continuously decreases to zero during the evolution process. Afterwards, the resulting variation equation is released by the FPEOP and this constraint returns to the inactive domain.

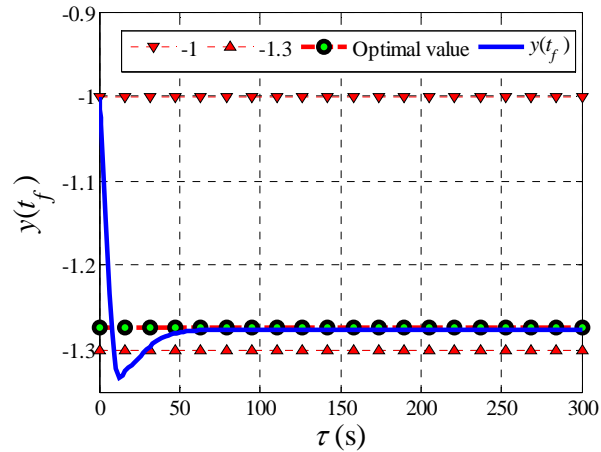

Fig. 7 The evolution profile of $y(t_f)$ and IEC bounds.

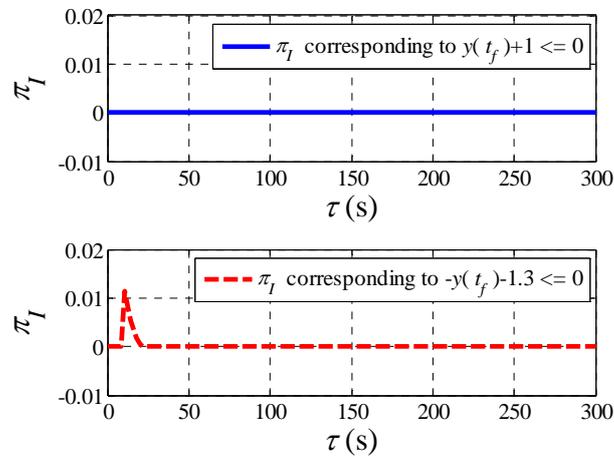

Fig. 8 The evolution profiles of KKT multipliers

## VI. Conclusion

The computation of Optimal Control Problems (OCPs) including terminal Inequality Constraint (IECs), under the frame of the Variation Evolving Method (VEM), is studied. To search the right evolution equations, the attributes of Equality Constraints (ECs) and IECs in a generalized optimization problem is traversed and the intrinsic relation to the multipliers is uncovered. The variation motion constraints arising from the active IECs are well distinguished, which preserves the feasibility of the variables and ensures the right evolution towards the optimal solution. The study gives an insight into the reasonable treatment of IECs with the VEM, and is foundational for the further study on OCPs with infinite-dimensional inequality path constraints. Actually the numerical soft barrier proposed to eliminate the numerical error hints it may turn the infeasible solution that violates the terminal IECs to be feasible, and this motivates us to further develop the VEM that is valid within the infeasible solution domain in the future.